\documentstyle[aps,epsf,twocolumn]{revtex}

\newcommand{\be}{\begin{equation}}
\newcommand{\ee}{\end{equation}}

\title{{\bf\Large Recognition of an organism from fragments of its complete 
genome\thanks{ Partially supported by the Australian Research Council grant A10024117, the
HKRGC Earmark Grant CUHK 4215/99P and a QUT postdoctoral fellowship. 
Email address of authors: v.anh@qut.edu.au (V.V. Anh), kslau@math.cuhk.edu.hk (K.S. Lau), 
yuzg@hotmail.com or z.yu@qut.edu.au (Z.G. Yu)}
}}
\author{V.V. Anh$^1$, K.S. Lau$^2$ and Z.G. Yu$^{1,3}$\thanks{Corresponding 
author.} \\
{\small $^1$Centre in Statistical Science and Industrial Mathematics,
Queensland University}\\
{\small of Technology, GPO Box 2434, Brisbane, Q4001, Australia}\\
{\small $^2$Department of Mathematics, Chinese University of Hong Kong,
Shatin, Hong Kong}\\
{\small $^3$Department of Mathematics, Xiangtan University, Hunan 411105,
P.R. China. }}

\begin{document}
\maketitle

\begin{abstract}
{\bf Abstract--} This paper considers the problem of matching fragment to 
organism using its
complete genome. Our method is based on the probability measure
representation of a genome. We first demonstrate that these probability
measures can be modelled as recurrent iterated function systems (RIFS)
consisting of four contractive similarities. Our hypothesis is that the
multifractal characteristic of the probability measure of a complete
genome, as captured by the RIFS, is preserved in its reasonably long
fragments. We compute the RIFS of fragments of various lengths and random
starting points, and compare with that of the original sequence for
recognition using the Euclidean distance. A demonstration on five randomly
selected organisms supports the above hypothesis.\newline
\newline
{\bf PACS} number(s): 87.14.Gg, 87.10+e, 47.53+n \newline
\end{abstract}

\maketitle

\section{Introduction}

\ \ \ \ The DNA sequences of complete genomes provide essential information for
understanding gene functions and evolution. A large number of these DNA
sequences is currently available in public databases such as Genbank at
ftp://ncbi.nlm.nih.gov/genbank/genomes/ or KEGG at
http://www.genome.ad.jp/kegg/java/org\_list.html). A great challenge of DNA
analysis is to determine the intrinsic patterns contained in these sequences
which are formed by four basic nucleotides, namely, adenine ($a$), cytosine (%
$c$), guanine ($g$) and thymine ($t$).

Some significant contribution results have been obtained for
the long-range correlation in DNA sequences
[1-16]. Li {\it et al.} \cite{li} found that the spectral density of a
DNA sequence containing mostly introns shows $1/f^{\beta }$ behaviour, which
indicates the presence of long-range correlation when $0<\beta <1$. The
correlation properties of coding and noncoding DNA sequences were first
studied by Peng {\it et al.} \cite{peng} in their fractal landscape or
DNA walk model. The DNA walk \cite{peng} was defined as that the walker steps
``up'' if a pyrimidine ($c$ or $t$) occurs at position $i$ along the DNA
chain, while the walker steps ``down'' if a purine ($a$ or $g$) occurs at
position $i$. Peng {\it et al.} \cite{peng} discovered that there exists
long-range correlation in noncoding DNA sequences while the coding sequences
correspond to a regular random walk. By undertaking a more detailed
analysis, Chatzidimitriou{\it \ et al.} \cite{CDL93} concluded that both
coding and noncoding sequences exhibit long-range correlation. A subsequent
work by Prabhu and Claverie \cite{PC92} also substantially corroborates
these results. If one considers more details by distinguishing $c$ from $t$
in pyrimidine, and $a$ from $g$ in purine (such as two or three-dimensional
DNA walk models \cite{luo} and maps given by Yu and Chen \cite{YC}), then the
presence of base correlation has been found even in coding sequences. On the
other hand, Buldyrev {\it et al.} \cite{Bul95} showed that long-range
correlation appears mainly in noncoding DNA using all the DNA sequences
available. Based on equal-symbol correlation, Voss \cite{voss} showed a
power law behaviour for the sequences studied regardless of the proportion
of intron contents. These studies add to the controversy about the possible
presence of correlation in the entire DNA or only in the noncoding DNA. From
a different angle, fractal analysis 
 has proven useful in revealing complex patterns in natural objects.
Berthelsen {\it et al.} \cite{bgs92} considered the global fractal
dimensions of human DNA sequences treated as pseudorandom walks.

In the above studies, the authors only considered  short or long DNA
segments. Since the first complete genome of the free-living bacterium {\it %
Mycoplasma genitalium} was sequenced in 1995 \cite{Fraser}, an
ever-growing number of complete genomes has been deposited in public
databases. The availability of complete genomes induces the possibility to
establish some global properties of these sequences. Vieira \cite{Vie99}
carried out a low-frequency analysis of the complete DNA of 13 microbial
genomes and showed that their fractal behaviour does not always prevail
through the entire chain and the autocorrelation functions have a rich
variety of behaviours including the presence of anti-persistence. Yu and Wang%
\cite{YW99} proposed a time series model of coding sequences in
complete genomes. For fuller details on the number, size and ordering of
genes along the chromosome, one can refer to Part 5 of Lewin \cite{Lew97}.
One may ignore the composition of the four kinds of bases in coding and
noncoding segments and only consider the global structure of the complete
genomes or long DNA sequences. Provata and Almirantis \cite{PY}
proposed a fractal Cantor pattern of DNA. They mapped coding segments to
filled regions and noncoding segments to empty regions of a random Cantor
set and then calculated the fractal dimension of this set. They found that
the coding/noncoding partition in DNA sequences of lower organisms is
homogeneous-like, while in the higher eucariotes the partition is fractal.
This result doesn't seem refined enough to distinguish bacteria because the
fractal dimensions of bacteria computed \cite{PY} are all the same. The
classification and evolution relationship of bacteria is one of the most
important problems in DNA research. Yu and Anh \cite{YA00} proposed a
time series model based on the global structure of the complete genome and
considered three kinds of length sequences. After calculating the
correlation dimensions and Hurst exponents, it was found that one can get
more information from this model than that of fractal Cantor pattern. Some
results on the classification and evolution relationship of bacteria were
found \cite{YA00}. The correlation property of these length sequences has
been discussed \cite{YAW00}. The multifractal analysis for these length sequences
was done in \cite{YAL01PA}.

Although statistical analysis performed directly on DNA sequences has
yielded some success, there has been some indication that this method is not
powerful enough to amplify the difference between a DNA sequence and a
random sequence as well as to distinguish DNA sequences themselves in more
details \cite{hlz98}. One needs more powerful global and visual methods.
For this purpose, Hao {\it et al.} \cite{hlz98} proposed a visualisation
method based on counting and coarse-graining the frequency of appearance of
substrings with a given length. They called it the {\it portrait} of an organism.
They found that there exist some fractal patterns in the portraits which are
induced by avoiding and under-represented strings. The fractal dimension of
the limit set of portraits was also discussed \cite{yhxc99,hxyc99}. There are
other graphical methods of sequence patterns, such as chaos game
representation \cite{Jef90,Gold93}.

Yu \emph{et al.} \cite{YAL01} introduced a representation of a DNA sequence by a
probability measure of $k$-strings derived from the sequence. This
probability measure is in fact the histogram of the events formed by all the 
$k$-strings in a dictionary ordering. It was found \cite{YAL01} that these
probability measures display a distinct multifractal behaviour characterised
by their generalised R\'{e}nyi dimensions (instead of a single fractal
dimension as in the case of self-similar processes). Furthermore, the
corresponding $C_{q}$ curves (defined in \cite{can00}) of these generalised dimensions of all bacteria
resemble classical phase transition at a critical point, while the
``analogous'' phase transitions (defined in \cite{can00}) of chromosomes of nonbacteria exhibit the
shape of double-peaked specific heat function. These patterns led to a
meaningful grouping of archaebacteria, eubacteria and eukaryote. Anh \emph{%
et al.} \cite{ALY} took a further step in providing a theory to characterise the
multifractality of the probability measures of the complete genomes. In
particular, the resulting parametric models fit extremely well the $D_{q}$
curves of the generalised dimensions and the corresponding $K_{q}$ curves of
the above probability measures of the complete genomes.

A conclusion of the work reported in Yu \emph{et al.} \cite{YAL01} and Anh \emph{et
al.} \cite{ALY} is that the histogram of the $k$-strings of the complete genome
provides a good representation of the genome and that these probability
measures are multifractal. This multifractality is, in most cases studied,
characteristic of the DNA sequences, hence can be used for their
classification.

In this paper, we consider the problem of recognition of an organism based
on fragments of their DNA sequences. The identification of the organisms in
a culture commonly relies on their molecular identity markers such as the
genes that code for ribosomal RNA. However, it is usual that most fragments
lack the marker, ``making the task of matching fragment to organism akin to
reconstructing a document that has been shredded'' (M. Leslie, ``Tales of
the sea'', \textit{New Scientist}, 27 January 2001). A well-known method to
tackle the task is the random shotgun sequencing method, which scans the
sequences of all fragments looking for overlaps to be able to piece the
fragments together. It is obvious that this technique is extremely
time-consuming and many crucial fragments may be missing.

This paper will provide a different method to approach this problem. Our
starting point is the probability measure of the $k$-strings and its
multifractality. We model this multifractality using a recurrent iterated
function system (\cite{BEH89,LN99}) consisting of
four contractive similarities (to be described in Section IV). This branching
number of four is a natural consequence of the four basic elements ($%
a,c,g,t) $ of the DNA sequences. Each of these RIFS is specified by a matrix
of incidence probabilities ${\bf P}=( p_{ij}) ,$ $i,j=1,...,4,$
with $p_{i1}+p_{i2}+p_{i3}+p_{i4}=1$ for $i=1,...,$ $4.$ It is our
hypothesis that, for reasonably-long fragments, the multifractal
characteristic of the measure of a complete genome as captured by the matrix 
$\mathbf{P}$ is preserved in the fragments. We thus represent each fragment
by a vector $( 
{\frac14}%
\left( p_{11}+p_{21}+p_{31}+p_{41}\right)$ ,
${\frac14}%
\left( p_{12}+p_{22}+p_{32}+p_{42}\right)$ ,
${\frac14}%
\left( \,p_{13}+p_{23}+p_{33}+p_{43} \right) $ in ${\bf R}%
_{+}^{3}. $ We will see that, for fragments of lengths longer than 1/20 of
the original sequence and with random starting points, these vectors are
very close, using the Euclidean distance, to the vector of the complete
sequence.

We will demonstrate the technique on five organisms, namely, {\it A. fulgidus, B.
burgdorferi, C. trachomatis, E. coli} and {\it M. genitalium}. As remarked in Yu 
\emph{et al.} \cite{YAL01}, substrings of length $k=6$ are sufficient to represent
DNA sequences. For each organism, we compute the histograms for the
6-strings of its complete genome, and 4 cases of fragments of lengths 1/4,
1/8, 1/15 and 1/20 of the complete sequence. The starting position of each
fragment is chosen randomly. The RIFS of the complete genome and each of the
fragments are computed next. The numerical results are reported in Section
V. Some conclusions will be drawn in Section VI.

\begin{figure}[tbp]
\centerline{\epsfxsize=8cm \epsfbox{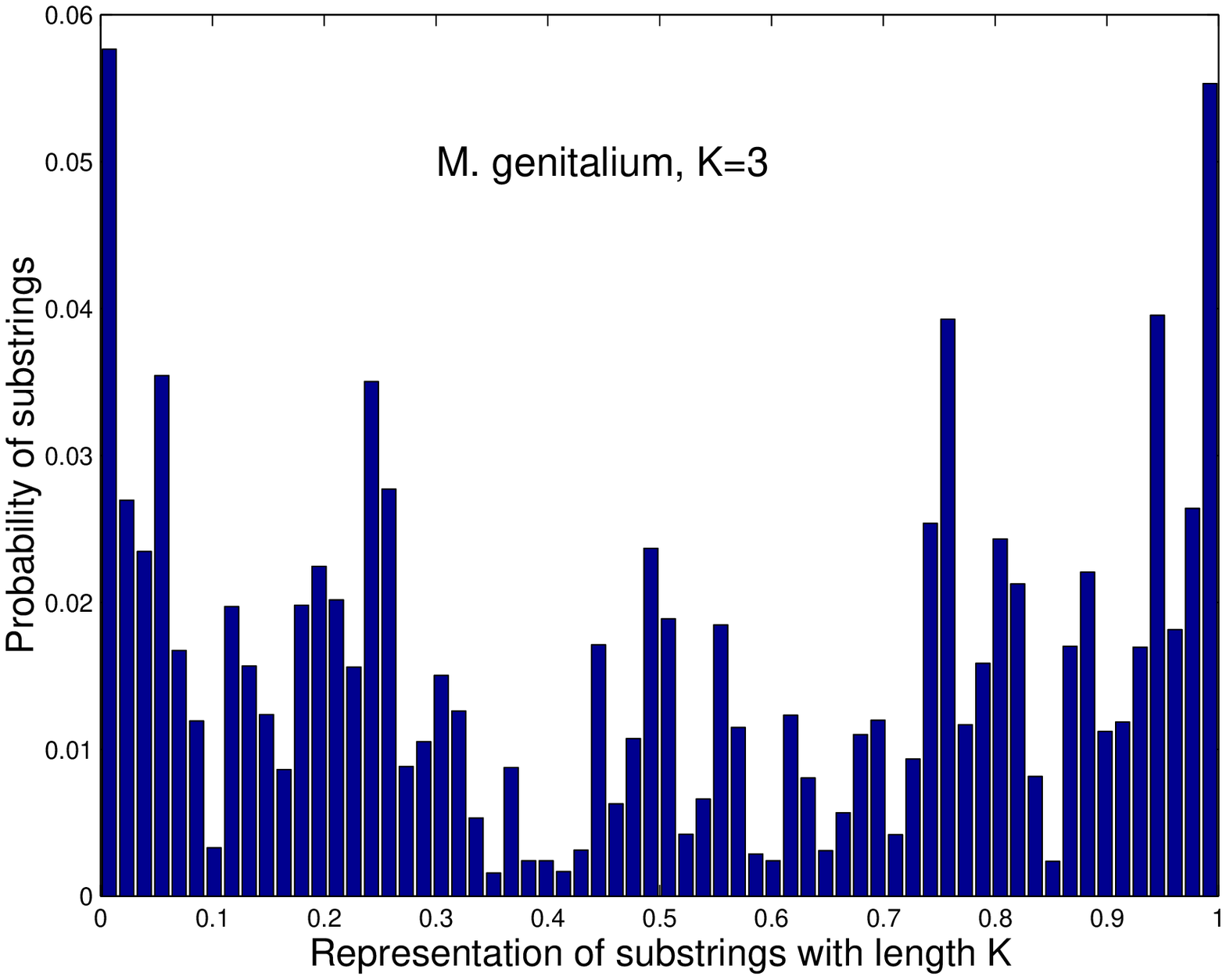}
\epsfxsize=8cm \epsfbox{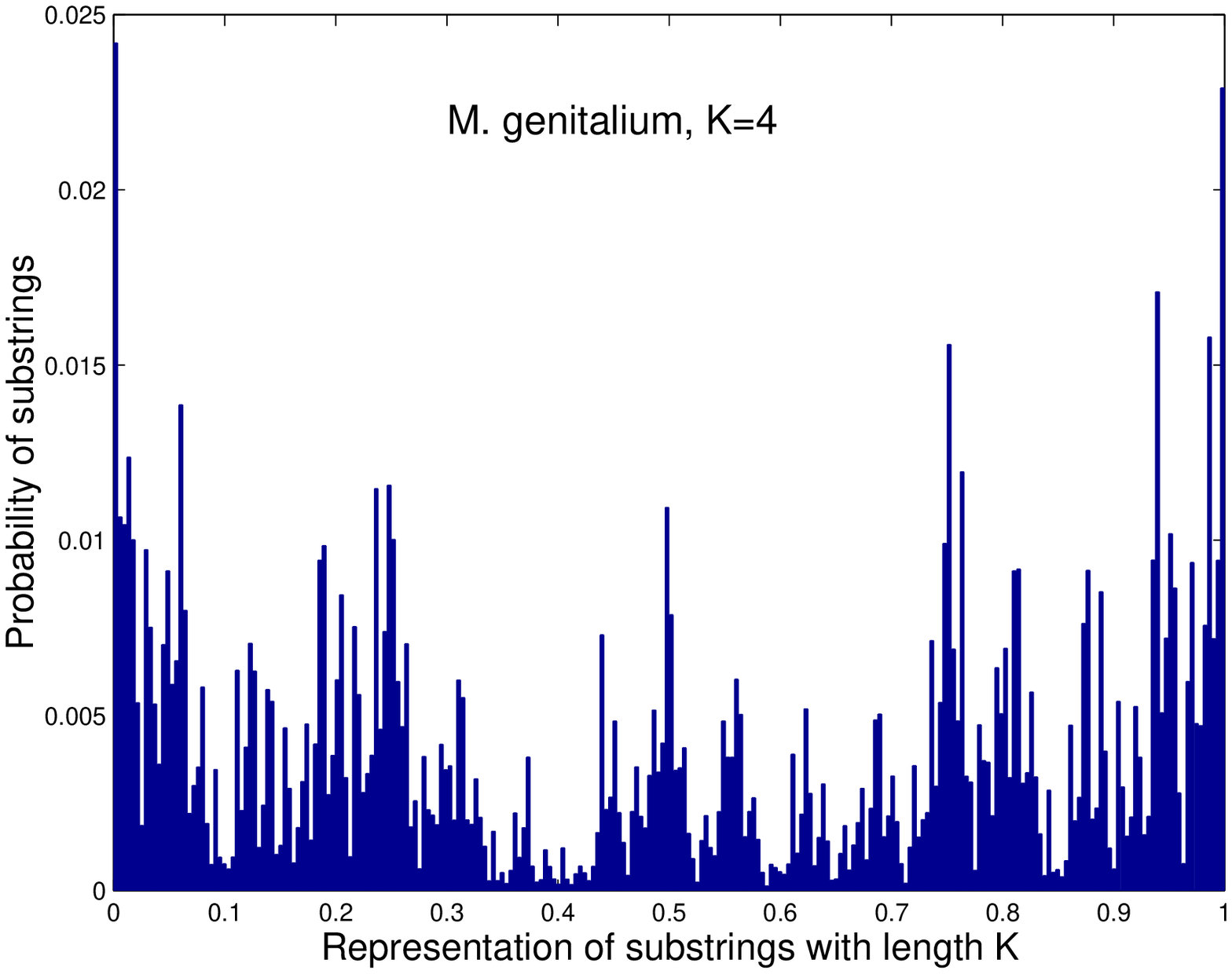}} 
\centerline{\epsfxsize=8cm \epsfbox{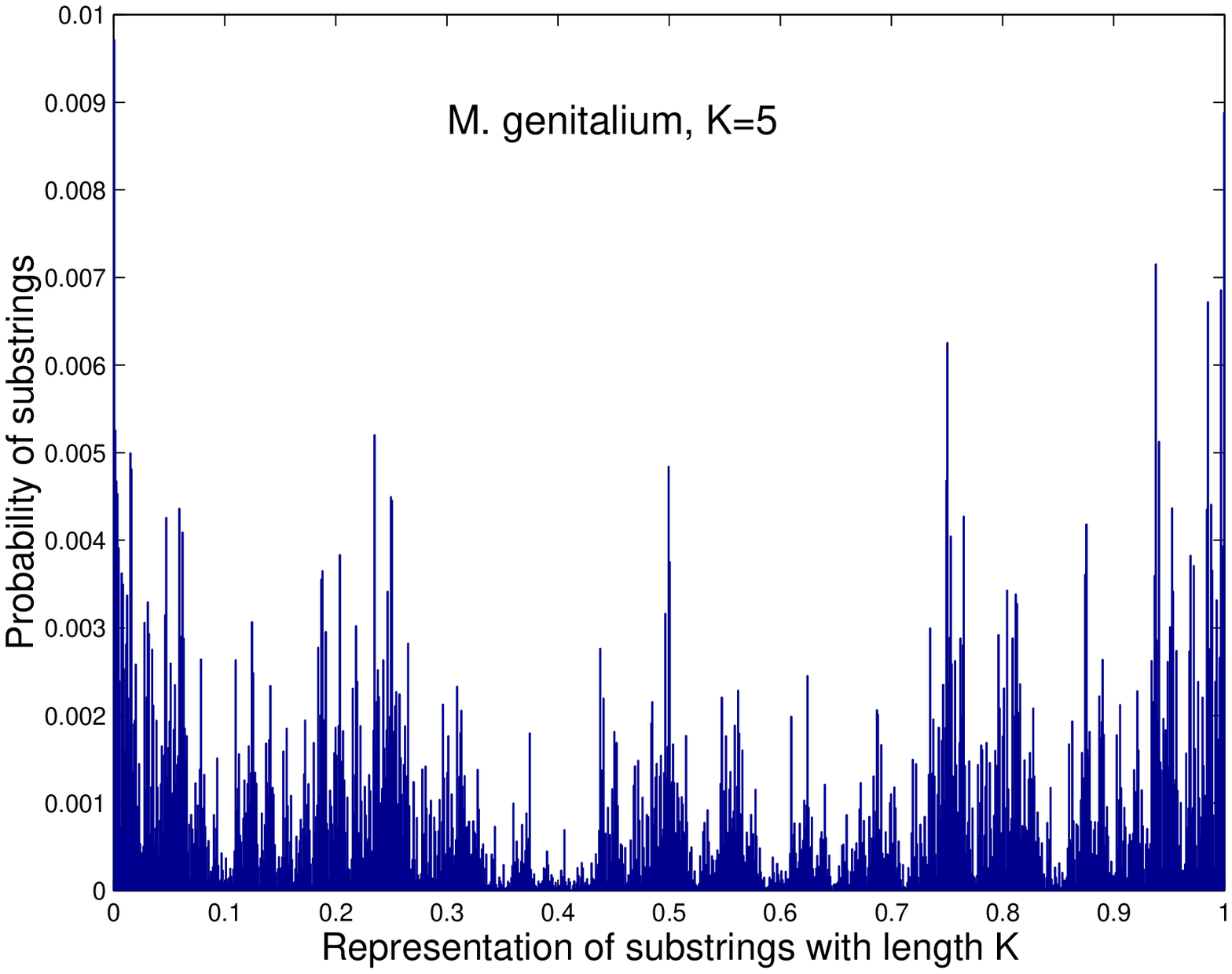}
\epsfxsize=8cm \epsfbox{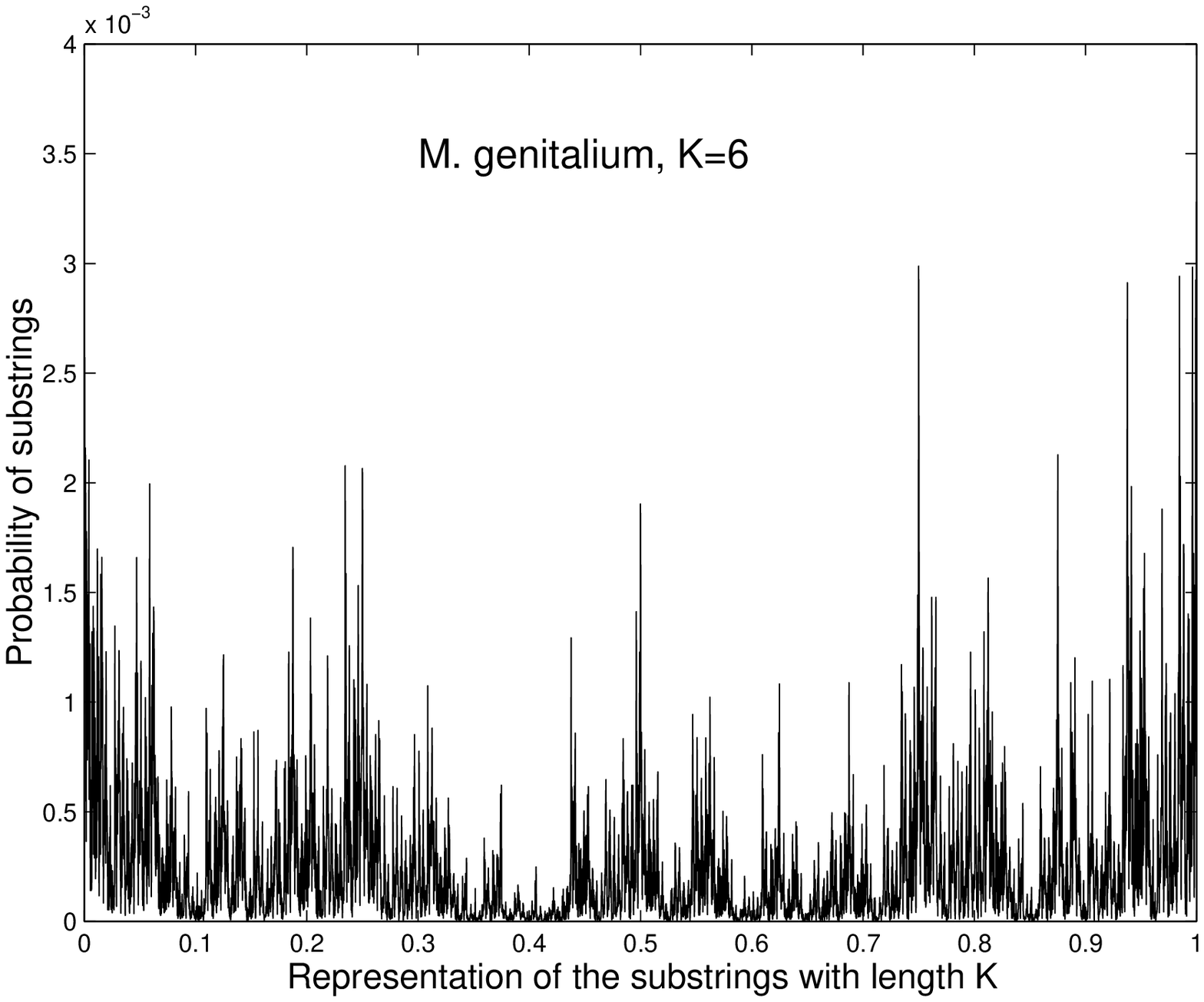}} 
\caption{{\protect\footnotesize Histograms of substrings with different
lengths}}
\label{mgencd}
\end{figure}

\section{Measure representation of complete genomes}

\ \ \ \ We first outline the method of Yu {\it et al.} \cite{YAL01} in deriving the
measure representation of a DNA sequence.\emph{\ }We call any string made up
of $k$ letters from the set $\{g,c,a,t\}$ a $k$-string. For a given $k$
there are in total $4^{k}$ different $k$-strings. In order to count the
number of each kind of $k$-strings in a given DNA sequence, $4^{k}$ counters
are needed. We divide the interval $[0,1)$ into $4^{k}$ disjoint
subintervals, and use each subinterval to represent a counter. Letting $%
s=s_{1}\cdots s_{k},\,s_{i}\in \{a,c,g,t\},i=1,\cdots ,k,$ be a substring
with length $k$, we define 
\begin{equation}
x_{l}(s)=\sum_{i=1}^{k}\frac{x_{i}}{4^{i}},  \label{2.1}
\end{equation}
where 
\begin{equation}
x_{i}=\left\{ 
\begin{array}{l}
0,\ \ \ \mbox{if}\ s_{i}=a, \\ 
1,\ \ \ \mbox{if}\ s_{i}=c, \\ 
2,\ \ \ \mbox{if}\ s_{i}=g, \\ 
3,\ \ \ \mbox{if}\ s_{i}=t,
\end{array}
\right.  \label{2.2}
\end{equation}
and 
\begin{equation}
x_{r}(s)=x_{l}(s)+\frac{1}{4^{k}}.  \label{2.3}
\end{equation}
We then use the subinterval $[x_{l}(s),x_{r}(s))$ to represent substring $s$%
. Let $N(s)$ be the times of substring $s$ appearing in the complete genome.
If the number of bases in the complete genome is $L$, we define 
\begin{equation}
F(s)=N(s)/(L-k+1)  \label{2.4}
\end{equation}
to be the frequency of substring $s$. It follows that $\sum_{\{s\}}F(s)=1$.
We can now view $F(s)$ as a function of $x$ and define a measure $\mu _{k}$
on $[0,1)$ by 
$$\mu _{k}\left( x\right) =Y_{k}\left( x\right) dx,
$$
where 
\begin{equation}
Y_{k}(x)=4^{k}F_{k}(s),\ \ \ \ x\in \lbrack x_{l}(s),x_{r}(s)).  \label{2.5}
\end{equation}
We then have $\mu _{k}\left( [0,1)\right) =1$ and $\mu _{k}\left(
[x_{l}(s),x_{r}(s))\right) =F_{k}(s).$ We call $\mu _{k}\left( x\right) $
the \textit{\ measure representation} of an organism. As an example, the
measure representation of \textit{M. genitalium} for $k=3,...,6$ is given
in FIG. \ref{mgencd}. A fractal-like behaviour is apparent in the measures.

{\bf Remark:}
{\it The ordering of $a,c,g,t$ in (\ref{2.2}) follows the natural dictionary
ordering of $k$-strings in the one-dimensional space. A different ordering
of $a,c,g,t$ would change the nature of the correlations of the measure. 
But in our case, a different ordering of $a,c,g,t$ in Eq. (\ref{2.2}) gives
the same multifractal spectrum ($D_q$ curve which will be defined in the next section)
 when the absolute value of $q$
is relatively small (see FIG. 2 in \cite{YAL01}). Hence the multifractal
characteristic is independent of the ordering. In
the comparison of different organisms using the measure representation, once
the ordering of $a,c,g,t$ in (\ref{2.2}) is given, it is fixed for all
organisms \cite{YAL01}.}

\section{Multifractal analysis}

\ \ \ \ The most common algorithms of multifractal 
analysis
are the so-called {\it fixed-size box-counting algorithms} \cite{hjkps}.
 In the one-dimensional case, for a given measure $\mu $ with support $%
E\subset {\bf R}$, we consider the {\it partition sum} 
\begin{equation}
Z_{\epsilon }(q)=\sum_{\mu (B)\neq 0}[\mu (B)]^{q},
\end{equation}
$q\in {\bf R}$, where the sum runs over all  different nonempty boxes $B$
of a given side $\epsilon $ in a grid covering of the support $E$, that is, 
\begin{equation}
B=[k\epsilon ,(k+1)\epsilon \lbrack .
\end{equation}
The exponent $\tau (q)$ is defined by 
\begin{equation}
\tau (q)=\lim_{\epsilon \rightarrow 0}\frac{\log Z_{\epsilon }(q)}{\log
\epsilon }
\end{equation}
and the generalized fractal dimensions of the measure are defined as 
\begin{equation}
D_{q}=\tau (q)/(q-1),\ \ \mbox{for}\ q\neq 1,
\end{equation}
and 
\begin{equation}
D_{q}=\lim_{\epsilon \rightarrow 0} \frac{Z_{1,\epsilon}}{\log \epsilon },\ \ \mbox{for}\ q=1,
\end{equation}
where $Z_{1,\epsilon}=\sum_{\mu (B)\neq 0}\mu (B)\log \mu(B)$.
The generalized fractal dimensions are estimated through a
linear regression of 
\[
\frac{1}{q-1}\log Z_{\epsilon }(q)
\]
against $\log \epsilon $ for $q\neq 1$, and similarly through a linear
regression of $Z_{1,\epsilon}$
against $\log \epsilon $ for $q=1$.  $D_1$ is called  {\it information dimension} and $D_2$ is called
 {\it correlation dimension}. The $D_q$ of the positive values of $q$ give relevance
to the regions where the measure is large, i.e., to the $k$-strings with
high probability. The $D_q$ of the negative values of $q$ deal with the
structure and the properties of the most rarefied regions of the measure.

\section{IFS and RIFS models and the moment method for parameter estimation}

\ \ \ \ In this paper, we propose to model the measure defined in Section II
for a complete genome by a recurrent IFS. As we work with measures on
compact intervals, the theory of Section II is narrowed down to the
one-dimensional case (i.e. $d=1).$ Consider a system of contractive maps $%
S=\{S_{1},S_{2},\cdots ,S_{N}\}$. Let $E_{0}$ be a compact interval of $%
{\bf R}$, $E_{\sigma _{1}\sigma _{2}\cdots \sigma _{n}}=S_{\sigma
_{1}}\circ S_{\sigma _{2}}\circ \cdots \circ S_{\sigma _{n}}(E_{0})$ and 
$$
E_{n}=\cup _{\sigma _{1},\cdots ,\sigma _{n}\in \{1,2,\cdots ,N\}}E_{\sigma
_{1}\sigma _{2}\cdots \sigma _{n}}.
$$
Then $E=\cap _{n=1}^{\infty }E_{n}$ is the \textit{attractor} of the IFS.
Given a set of probabilities $p_{i}>0,\ \sum_{i=1}^{N}p_{i}=1$, we pick an $%
x_{0}\in E$ and define iteratively the sequence 
\begin{equation}
x_{n+1}=S_{\sigma _{n}}(x_{n}),\qquad \quad n=0,1,2,\cdots ,  \label{4.1}
\end{equation}
where the indices $\sigma _{n}$ are chosen randomly and independently from
the set $\{1,2,\cdots ,N\}$ with probabilities $P(\sigma _{n}=i)=p_{i}$.
Then every orbit $\{x_{n}\}$ is dense in the attractor $E$ \cite
{barnsley,vrscay}. For $n$ large enough, we can view the orbit $%
\{x_{0},x_{1},\cdots ,x_{n}\}$ as an approximation of $E$. This iterative
process is called a \textit{chaos game}.

Given a system of contractive maps $S=\{S_{1},S_{2},\cdots ,S_{N}\}$ on a
compact metric space $E^{\ast }$, we associate with these maps a matrix of
probabilities ${\bf P}=(p_{ij})$ such that $\sum_{j}p_{ij}=1,\
i=1,2,\cdots ,N$. Consider a random sequence generated by a chaos game: 
\begin{equation}
x_{n+1}=S_{\sigma _{n}}(x_{n}),\quad n=0,1,2,\cdots ,  \label{4.2}
\end{equation}
where $x_{0}$ is any starting point and $\sigma _{n}$ is chosen with a
probability that depends on the previous index $\sigma _{n-1}$: 
\begin{equation}
P(\sigma _{n+1}=i)=p_{\sigma _{n},i}.  \label{4.3}
\end{equation}
The choice of the indices $\sigma _{n}$ as prescribed by (\ref{4.3})
presents a fundamental difference between this iterative process and that
defined by (\ref{4.1}) of the usual chaos game. Then $(E^{\ast },S,{\bf P}%
)$ is called a \textit{recurrent IFS}. The flexibility of RIFS permits the
construction of more general sets and measures which do not have to exhibit
the strict self-similarity of IFS. This would offer a more suitable
framework to model fractal-like objects and measures in nature.

Let $\mu $ be the invariant measure on the attractor $E$ of an IFS or RIFS, $%
\chi _{B}$ the characteristic function for the Borel subset $B\subset E$;
then from the ergodic theorem for IFS or RIFS \cite{barnsley}, 
\begin{equation}
\mu (B)=\lim_{n\rightarrow \infty }[\frac{1}{n+1}\sum_{k=0}^{n}\chi
_{b}(x_{k})].  \label{4.4}
\end{equation}
In other words, $\mu (B)$ is the relative visitation frequency of $B$
during the chaos game. A histogram approximation of the invariant measure
may then be obtained by counting the number of visits made to each pixel on
the computer screen.

 The coefficients in the contractive maps and the probabilities in
the IFS or RIFS model are the parameters to be estimated for a given measure
which we want to simulate. Vrscay \cite{vrscay} introduced a moment method
to perform this task. If $\mu $ is the invariant measure and $E$ the
attractor of the IFS or RIFS in ${\bf R}$, the moments of $\mu $ are 
\begin{equation}
g_{i}=\int_{E}x^{i}d\mu ,\qquad g_{0}=\int_{E}d\mu =1.  \label{judy}
\end{equation}
If $S_{i}(x)=c_{i}x+d_{i},\ i=1,\cdots ,N$, then the following well-known
recursion relations hold for the IFS model: 
\begin{equation}
\lbrack 1-\sum_{i=1}^{N}p_{i}c_{i}^{n}]g_{n}=\sum_{j=1}^{n}\left( 
\begin{array}{l}
n \\ 
j
\end{array}
\right) g_{n-j}(\sum_{i=1}^{N}p_{i}c_{i}^{n-j}d_{i}^{j}).  \label{jugsh1}
\end{equation}
Thus, setting $g_{0}=1$, the moments $g_{n},\ n\geq 1$, may be computed
recursively from a knowledge of $g_{0},\cdots ,g_{n-1}$ \cite{vrscay}.

For the RIFS model, we have 
\begin{equation}
g_n=\sum_{j=1}^N g_n^{(j)},  \label{jushi2}
\end{equation}
where $g_n^{(j)},\ j=1,\cdots,N$, are given by the solution of the following
system of linear equations: 
\begin{eqnarray}
\sum_{j=1}^N(p_{ji}c_i^n-\delta_{ij})g_n^{(j)}&=&-\sum_{k=0}^{n-1} \left( 
\begin{array}{l}
n \\ 
k
\end{array}
\right) [\sum_{j=1}^Nc_i^k d_i^{n-k}p_{ji}g_k^{(j)}], \nonumber\\ 
& &i=1,\cdots,N,\ n\ge³1. 
\label{jushi3}
\end{eqnarray}
For $n=0$, we set $g_0^{(i)}=m_i$, where $m_i$ are given by the solution of
the linear equations 
\begin{equation}
\sum_{j=1}^Np_{ji}m_j=m_i,\quad i=1,2,\cdots,N,\quad 
g_0=\sum_{i=1}^Nm_i=1.  \label{jushi4}
\end{equation}

If we denote by $G_{k}$ the moments obtained directly from a given measure
using (\ref{judy}), and $g_{k}$ the formal expression of moments obtained
from (\ref{jugsh1}) for the IFS model or from (\ref{jushi2}-\ref{jushi4})
for the RIFS model, then through solving the optimal problem 
\begin{equation}
\min_{c_{i},d_{i},p_{i}\mbox{ or }p_{ij}}\sum_{k=1}^{n}(g_{k}-G_{k})^{2},%
\qquad \mbox{for some chosen}\ n,  \label{youhua}
\end{equation}
we can obtain the estimates of the parameters in the IFS or RIFS model.

From the measure representation of a complete genome, it is natural to
choose $N=4$ and 
\begin{eqnarray*}
& &S_{1}(x)=x/4,\ S_{2}(x)=x/4+1/4,\\
& &S_{3}(x)=x/4+1/2,\ S_{4}(x)=x/4+3/4
\end{eqnarray*}
in the IFS or RIFS model. Based on the estimated values of the
probabilities, we can use the chaos game to generate a histogram
approximation of the invariant measure of the IFS or RIFS, which then can be
compared with the given measure of the complete genome.

\begin{figure}[tbp]
\centerline{\epsfxsize=8cm \epsfbox{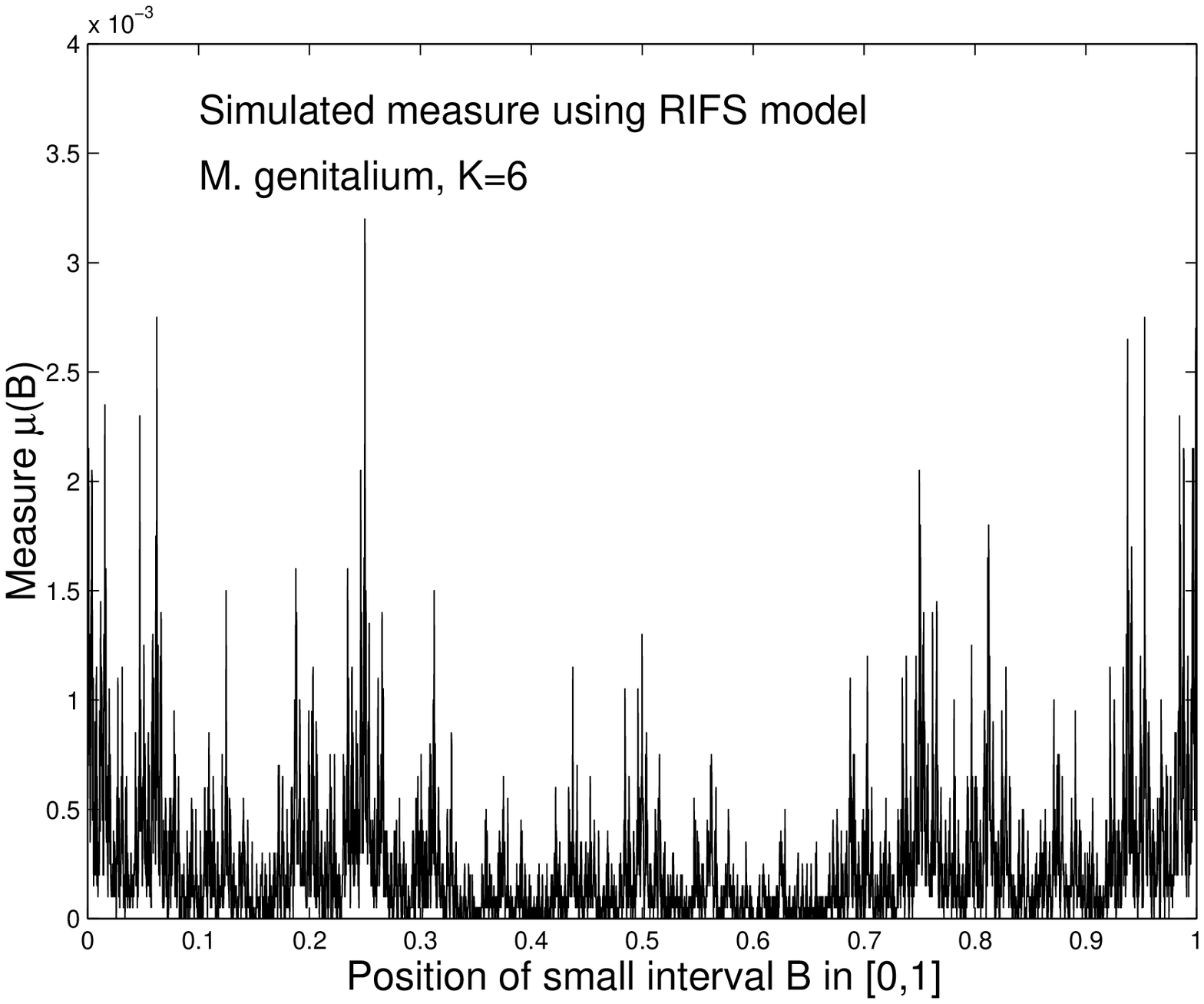}
\epsfxsize=8cm \epsfbox{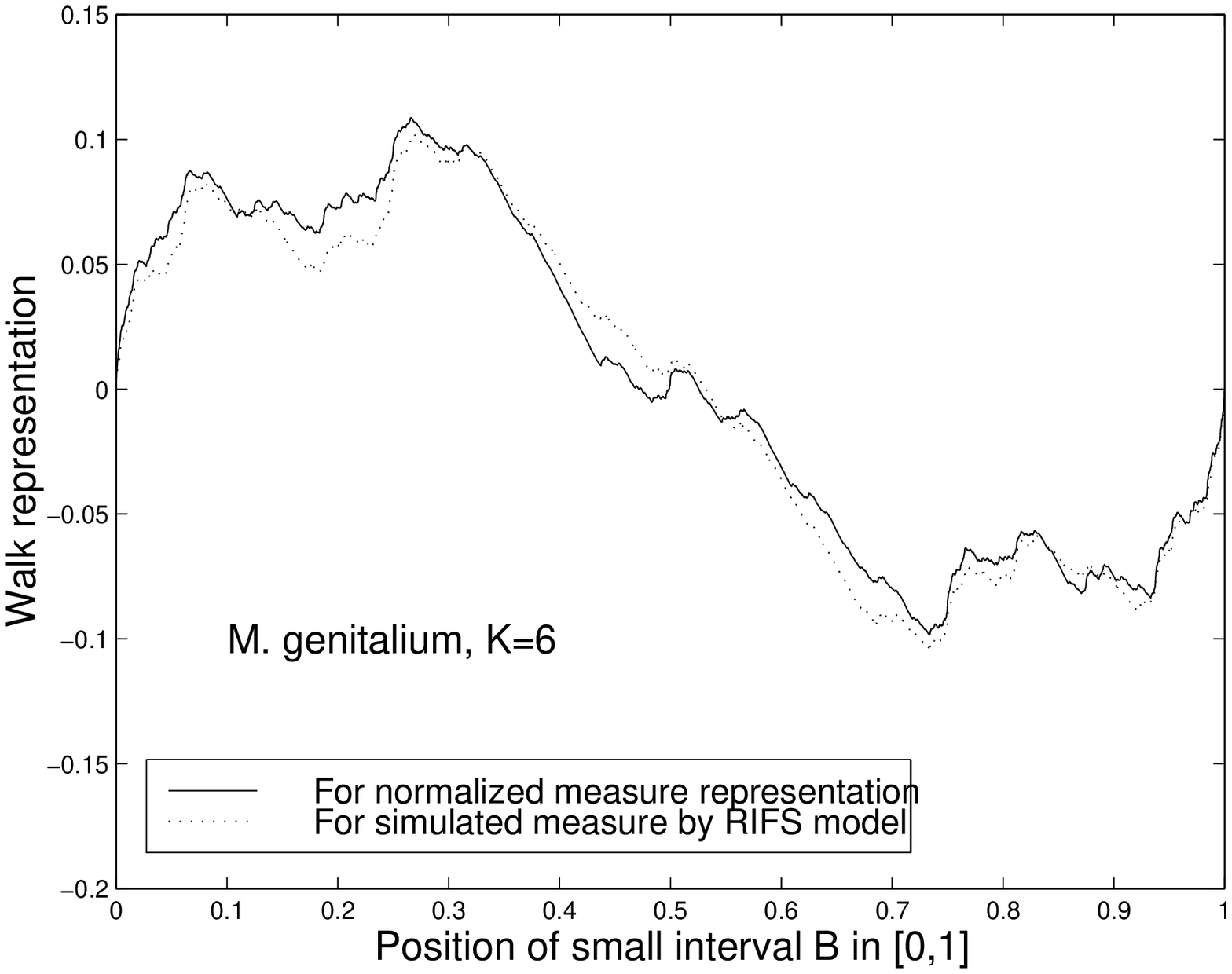}} 
\caption{{\protect\footnotesize {\bf Left):} Simulation of the measure 
representation (6-strings) of the whole genome of {\it M. genitalium} using the recurrent IFS model.
{\bf Right):} Walk comparison for measure representation (6-strings) of 
{\it M. genitalium} and
its RIFS simulation. }}
\label{mgensimu}
\end{figure}

\section{Application to the recognition problem}

\ \ \ \ The measure representations for a large number of complete genomes, as
described in Section II, were obtained in Yu \emph{et al. } \cite{YAL01}. It was found
that substrings with $k=6$ seem to provide a limiting measure that can be
used for the classification and recognition of DNA sequences. Hence we will
use 6-strings in this paper. We then estimated their IFS and RIFS models
using the moment method described in Section 4. The chaos game algorithm was
next performed to generate an orbit as in (\ref{4.1}) or (\ref{4.2}) with (%
\ref{4.3}). From these orbits, simulated approximations of the invariant
measures of IFS or RIFS were obtained via the ergodic theorem (\ref{4.4}).
In order to clarify how close the simulated measure is to the original
measure, we convert a measure to its walk representation: We denote by $%
\{t_{j},\ j=1,2,\cdots ,4^{k}\}$ the density of a measure and $t_{ave}$ its
average, then define the walk $T_{j}=\sum_{k=1}^{j}(t_{k}-t_{ave}),\
j=1,2,\cdots ,4^{k}$. The two walks of the given measure and the measure
generated by the chaos game of an IFS or RIFS are then plotted in the same
figure for comparison. We found that RIFS is a better model to simulate
complete genomes. We determine the "goodness" of the measure simulated from the RIFS model
relative to the original measure based on the following {\it relative
standard error} ({\it RSE})
$$ RSE=\frac{RMSE}{SE},$$
where
$$RMSE=\sqrt{\frac{1}{4^6}\sum_{j=1}^{4^6}(t_j-\hat{t}_j)^2},$$
and
$$SE=\sqrt{\frac{1}{4^6}\sum_{j=1}^{4^6}(t_j-t_{ave})^2},$$
$(t_j)_{j=1}^{4^6}$ and $(\hat{t}_j)_{j=1}^{4^6}$ being the densities of the original
measure and the RIFS simulated measure respectively. The goodness
of fit is indicated by the result $RSE \ < \ 1$.
For example, the RIFS simulation of 6-strings measure 
representation of {\it M. genitalium} is shown in the left figure of FIG. \ref{mgensimu},
and the walk of its original 6-strings measure representation
and that simulated from the corresponding 
RIFS are shown in the right figure of FIG. \ref{mgensimu}. 
For the whole genome, $RMSE=0.00020675$, $SE=0.0003207$ and 
$RSD=0.6447\ < \ 1$. It is seen that the RIFS
simulation fits the original measure very well.

\begin{figure}[tbp]
\centerline{\epsfxsize=8cm \epsfbox{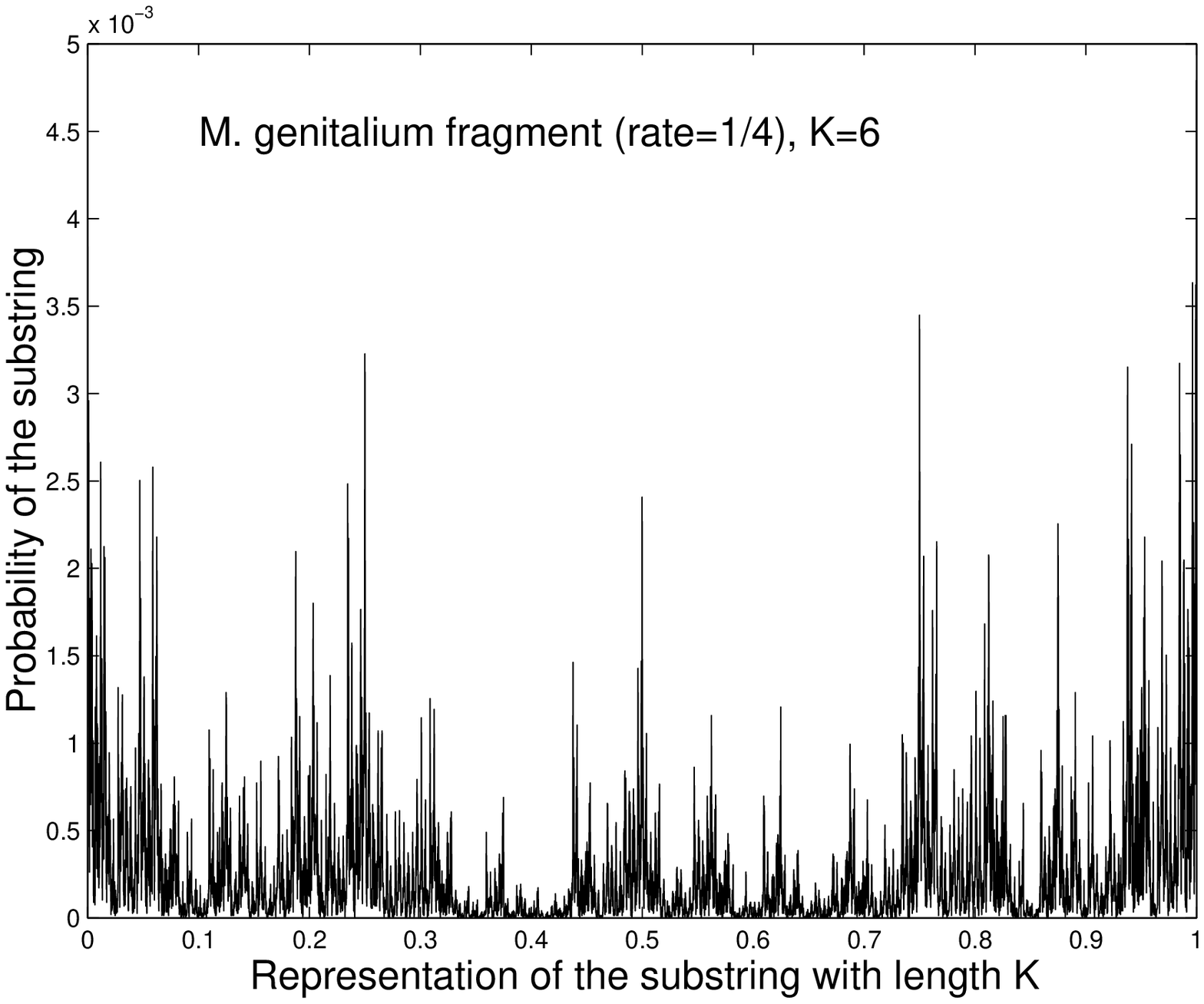}
\epsfxsize=8cm \epsfbox{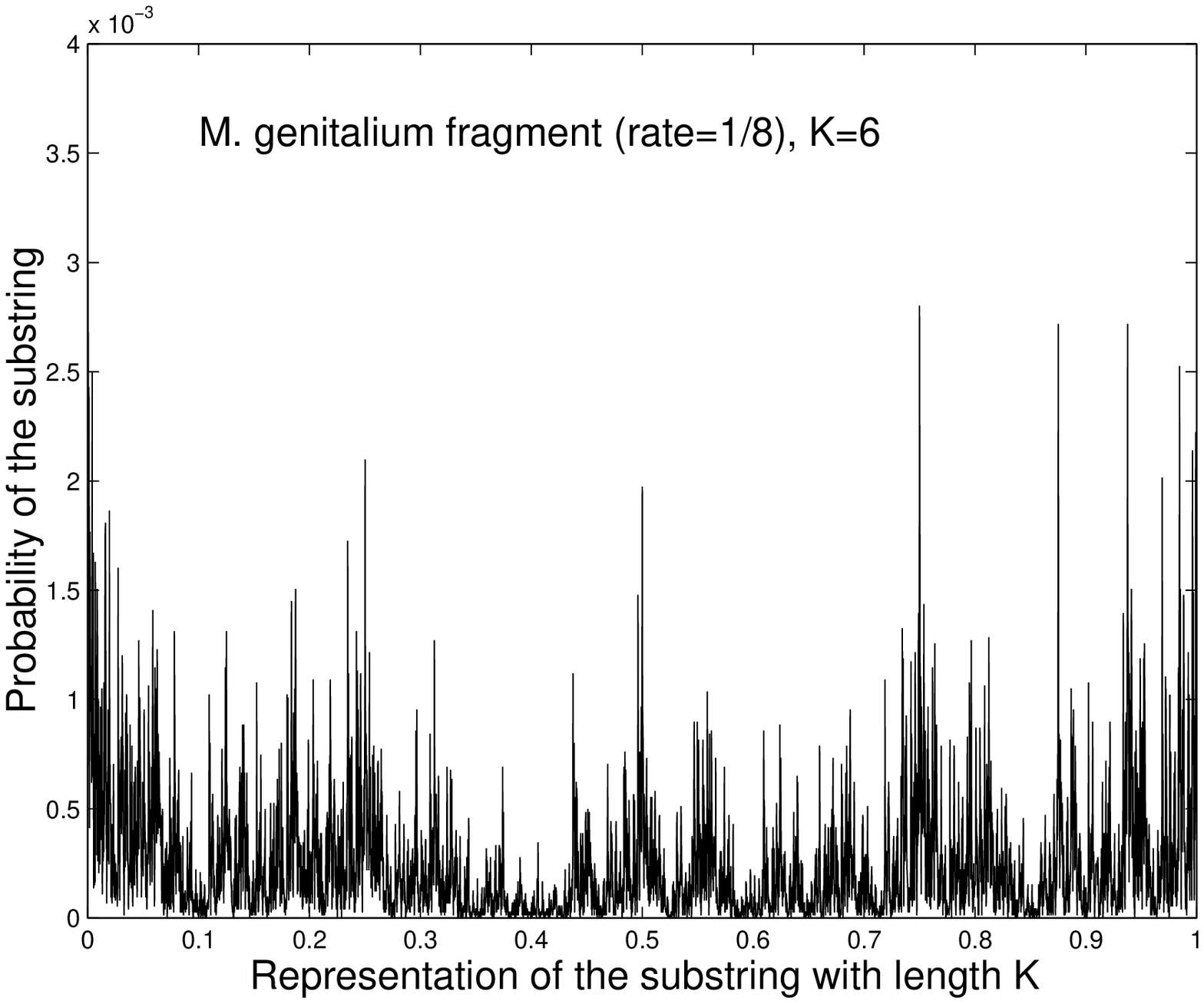}}
\centerline{\epsfxsize=8cm \epsfbox{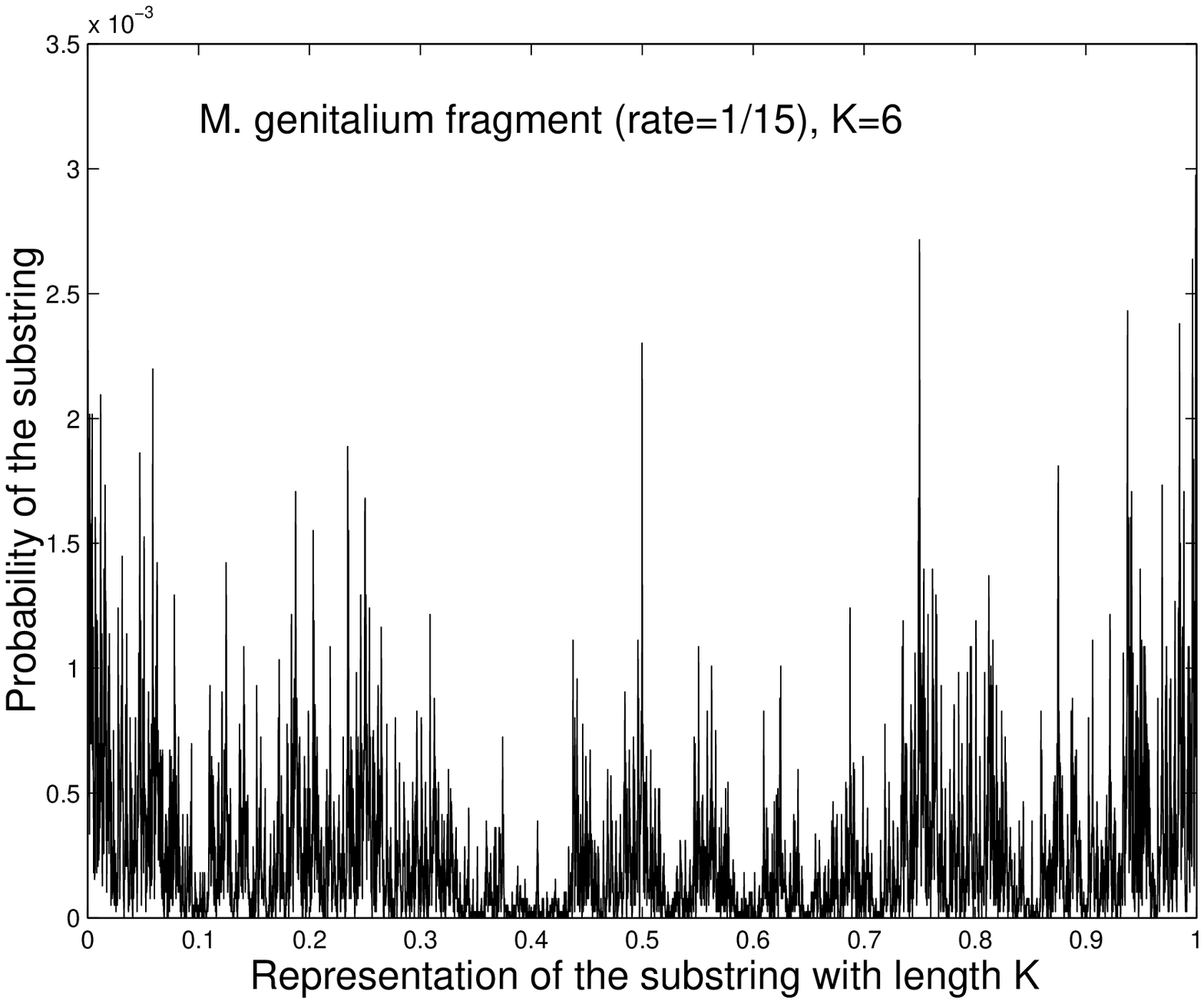}
\epsfxsize=8cm \epsfbox{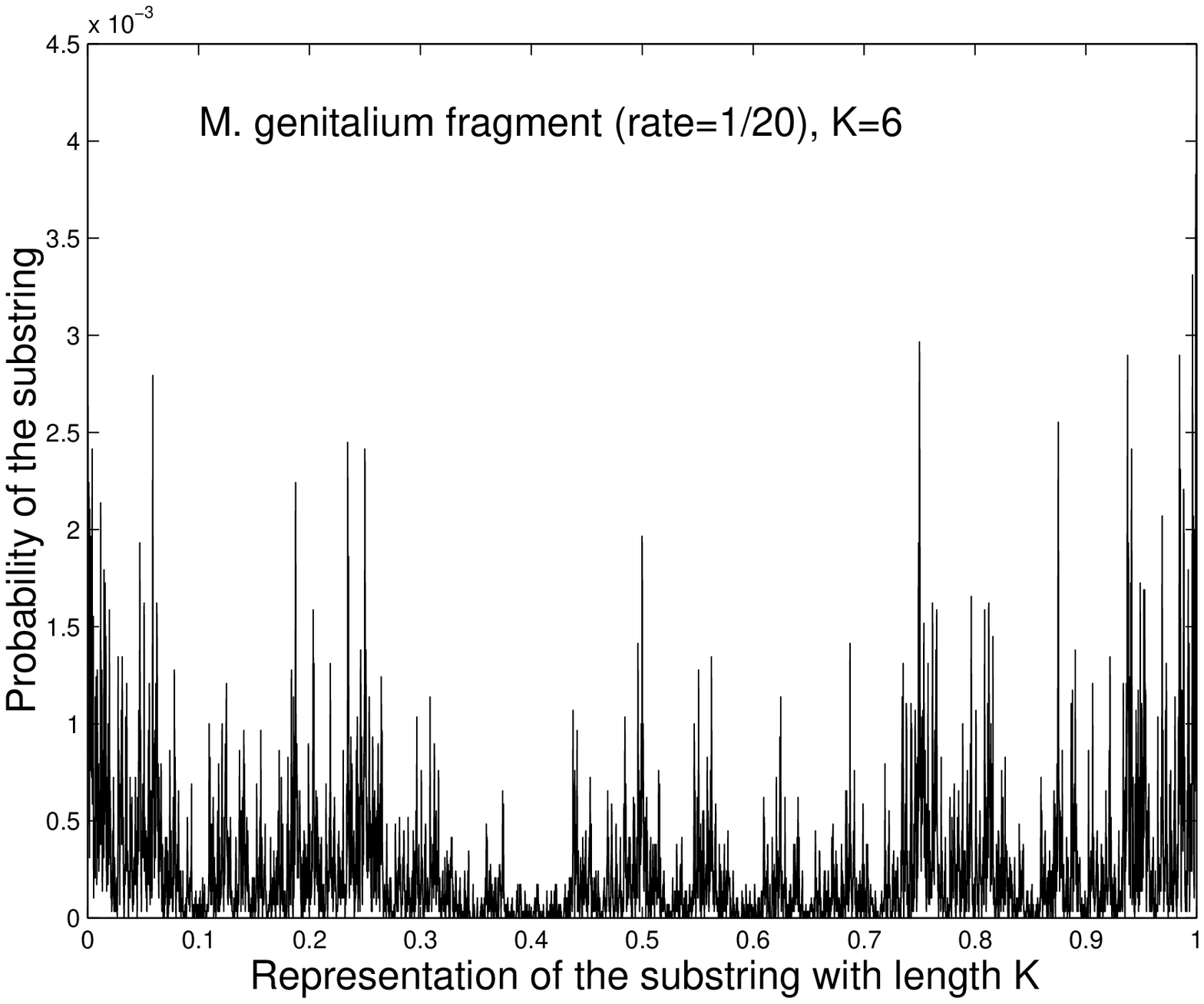}} 
\caption{{\protect\footnotesize Histograms of 6-substrings of fragments from
{\it M. genitalium} with different rates.}}
\label{mgenfragcd}
\end{figure}

\begin{figure}[tbp]
\centerline{\epsfxsize=8cm \epsfbox{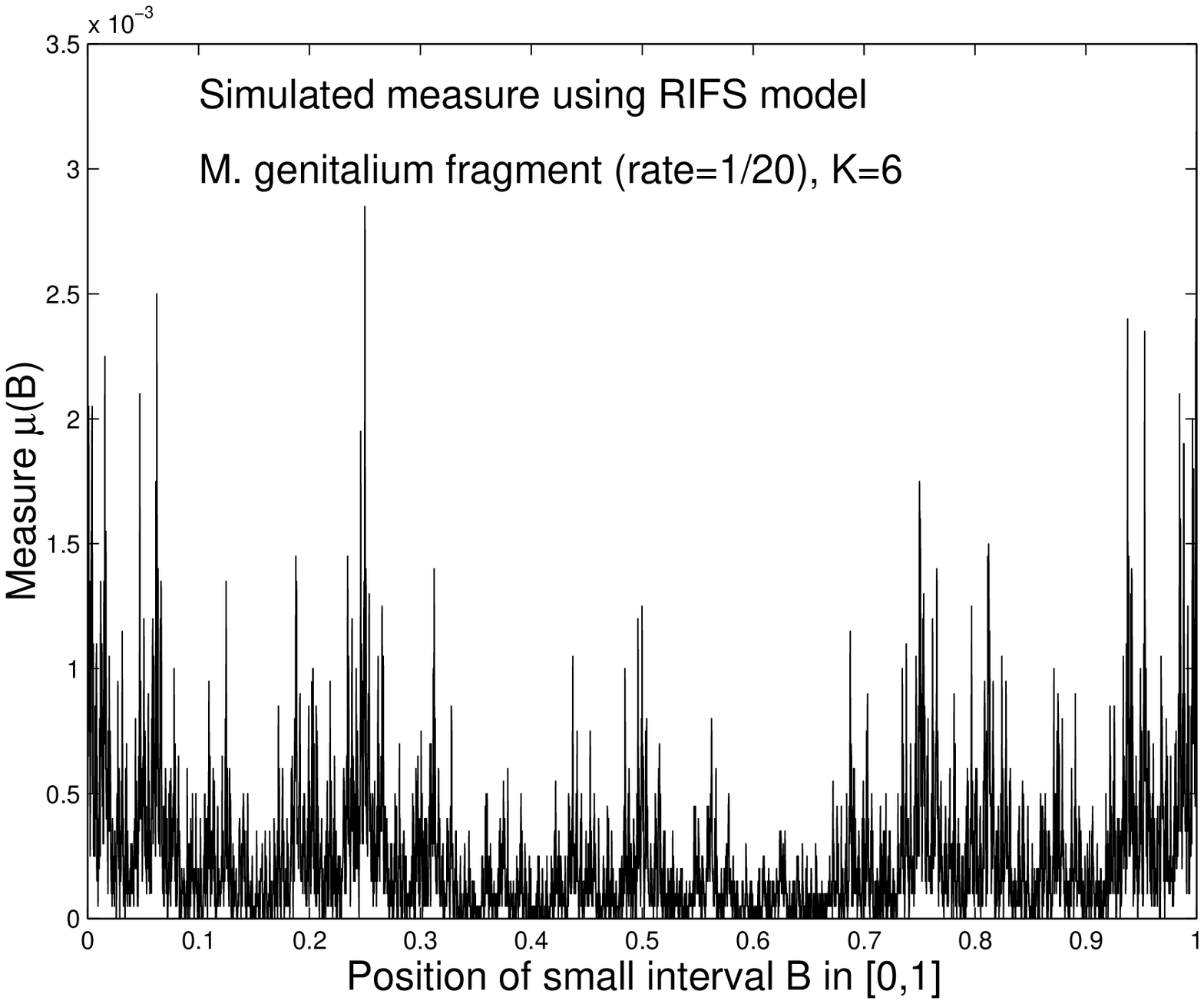}
\epsfxsize=8cm \epsfbox{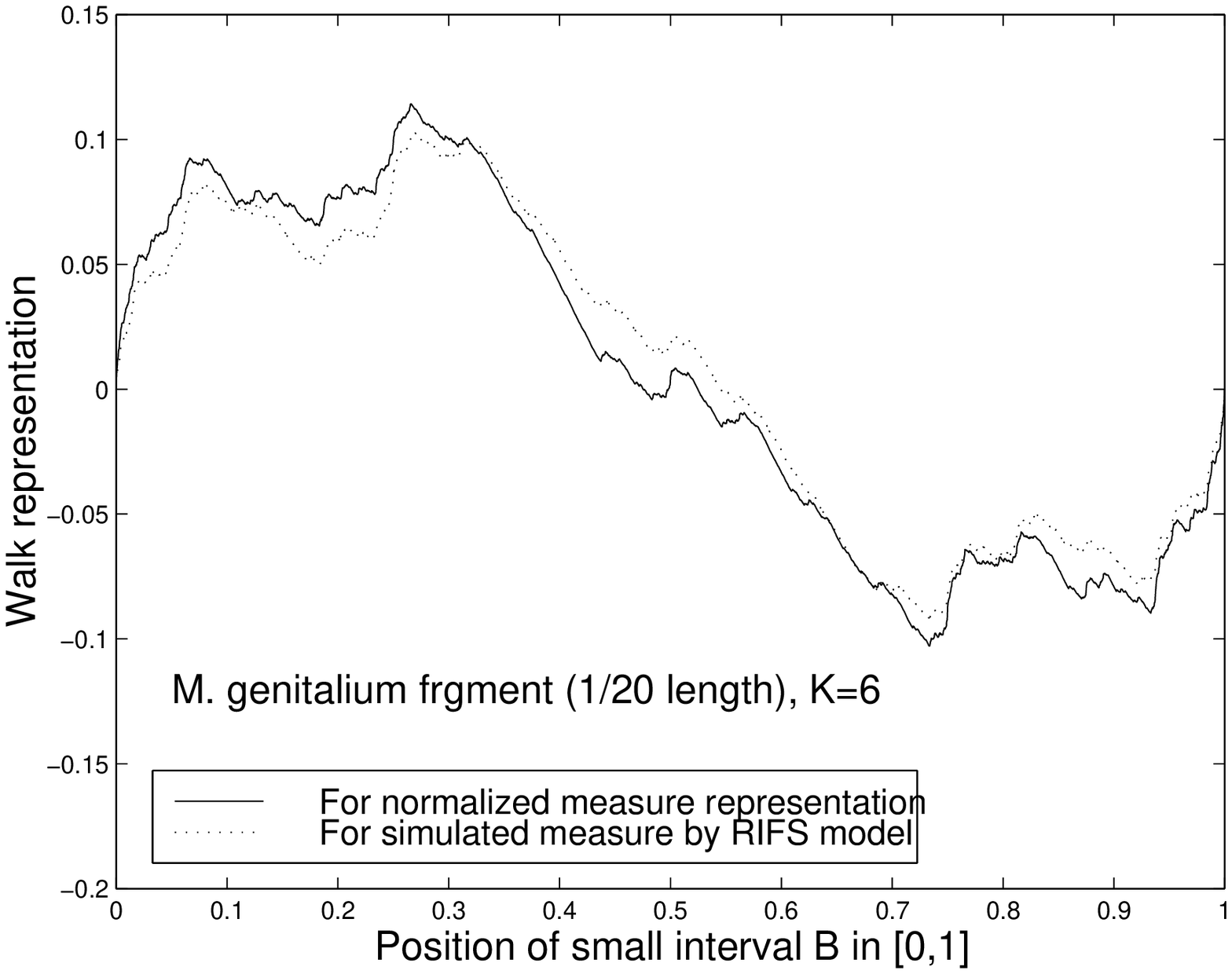}} 
\caption{{\protect\footnotesize {\bf Left):} Simulation of the measure 
representation (6-strings) of 1/20 fragment of {\it M. genitalium} using the 
recurrent IFS model.
{\bf Right):} Walk comparison for measure representation (6-strings) of 1/20 fragment 
of {\it M. genitalium} and
its RIFS simulation. }}
\label{mgenfragsimu}
\end{figure}

We next pick out five organisms (without any particular \emph{a priori }%
reason) from about 50 organisms whose complete genomes are currently
available. These are {\it A. fulgidus, B. burgdorferi, C. trachomatis, E. coli}
and {\it M. genitalium}. Fragments of different length rates ranging from 1/20 to 1/4
and with random starting points along the sequences were then selected. 
Here the length rate of a fragment means the length of this fragment divided by
the length of the genome of the same organism. For
example, the measure representations of different fragments of \textit{M.
genitalium }are shown in FIG. \ref{mgenfragcd}. The RIFS model for each of
these fragments was next estimated. 
We also show the RIFS simulation of the 6-strings measure representation of the 
1/20 fragment of {\it M. genitalium} in the left figure of FIG. \ref{mgenfragsimu}.
The walk of its original 6-strings measure representation and that of RIFS simulation
are shown in the right figure of FIG. \ref{mgenfragsimu}. For this fragment,
$RMSE=0.00023169$, $SE=0.00035475$ and 
$RSD=0.6531\ < \ 1$. Again, the RIFS simulation
fits the original measure of this fragment very well.

\begin{figure}[tbp]
\centerline{\epsfxsize=10cm \epsfbox{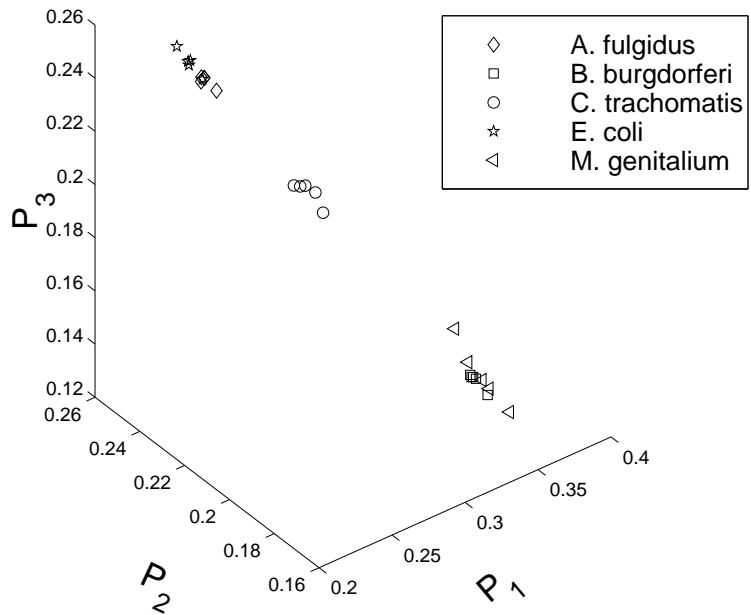}}
\caption{{\protect\footnotesize Vector representation ($P_1,P_2,P_3)$) of all fragments
from five organisms.}}
\label{plotp1p2p3}
\end{figure}

It should be noted that column $i$ in
the matrix $\mathbf{P}$ describes the activity of similarity $S_{i}$ in each
RIFS. To be able to represent each fragment on a three-dimensional plot, we
define 
\begin{equation}
\left\{ 
\begin{array}{l}
P_{1}=(p_{11}+p_{21}+p_{31}+p_{41})/4,\\
P_{2}=(p_{12}+p_{22}+p_{32}+p_{42})/4,\\
P_{3}=(p_{13}+p_{23}+p_{33}+p_{43})/4. 
\end{array} \right.
\label{5.1}
\end{equation}
Each fragment is then represented by the vector $\left(
P_{1},P_{2},P_{3}\right) .$ The values of these vectors are provided in
Table \ref{vectortab}, and the vectors are plotted in FIG. \ref{plotp1p2p3}. 
It is seen that the
vectors of the fragments from the same organism cluster together, and this
clustering holds for all selected lengths. This accuracy is uniform for all
five organisms randomly selected.

\begin{figure}[tbp]
\centerline{\epsfxsize=12cm \epsfbox{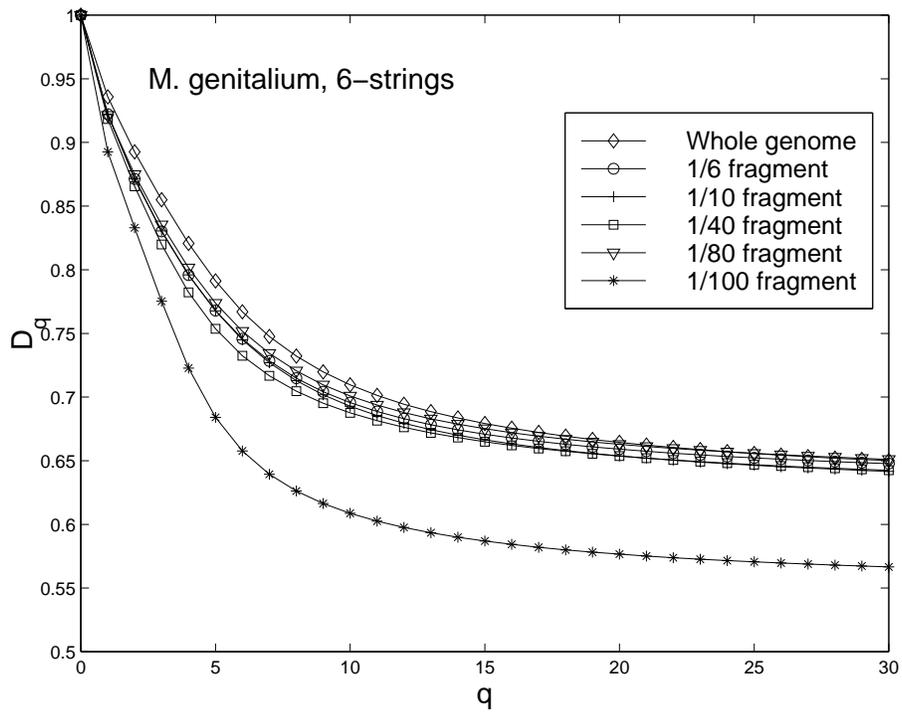}}
\caption{{\protect\footnotesize The dimension spectra of fragments from {\it M. genitalium}.}}
\label{mgenDq}
\end{figure}

In matching a fragment to organism, the $D_{q}$ curve, which depicts the
generalised dimension of the invariant measure as described in Section III,
can also be used. We computed these curves for the above five organisms at a
variety of length sizes, to 1/100th of the original sequence. The results
were reported for {\it M. genitalium} in FIG. \ref{mgenDq}. It is seen that this method
also performs very well. However, it suffers a drawback that many different
organisms seem to have the same or closely related $D_{q}$ curve. In this
sense, the method based on the RIFS has higher resolution in
distinguishing the genomes.  If necessary, the entire matrix $\mathbf{P}$ may
be used, instead of (\ref{5.1}), in this comparison. This would enhance the
matching, but will not be as economical as (\ref{5.1}). Yu {\it et al.} \cite{YALCjh}
used the entire matrix {\bf P} to define the distance between two organisms 
in higher dimensional space and then the evolutionary tree of more than 50 organisms
was constructed. The RIFS model can also be used to simulate the measure representation
of proteins based on the HP model \cite{YAL02}.

\section{Conclusion}

\ \ \ \ This paper provides a method for matching fragment to organism taking
advantage of the multifractal characteristic of the measure representation
of their genomes. It was demonstrated empirically that the underlying
mechanism of this multifractality can be captured by a recurrent IFS, whose
theory is well founded in the fractal geometry literature. Fast algorithms
for the computation of these RIFS and related quantities as well as tools
for comparison are available. The method seems to work reasonably well with
low computing cost. This fast and economical method can be performed at a
preliminary stage to cluster fragments before a more extensive method, such
as the random shotgun sequencing method as mentioned in the Introduction, is
decided to be brought in for higher accuracy.

\onecolumn

 \begin{table}[tbp]
\caption{Values of vector representation $(P_1,P_2,P_3)$ of fragments from the five organisms.}
\begin{center}
\begin{tabular}{|l|c|c|c|c|}
  Organism & Sequence &    $P_1$ & $P_2$ & $P_3$\\
 \hline
  & 1/4 fragment & 0.255114 &0.248454 &0.234208\\
  & 1/8 fragment& 0.257610 & 0.248891 & 0.232988 \\
A. fulgidus   & 1/15 fragment & 0.260611 & 0.245235 & 0.229882 \\
  & 1/20 fragment&  0.253536 & 0.247569 & 0.233501 \\
  & whole genome &0.257277 &0.248579 &0.233379 \\
\hline  
 &1/4 fragment & 0.305165 &0.160478 & 0.165485 \\
 & 1/8 fragment & 0.303635 & 0.160063 & 0.166952 \\
B. burgdorferi  & 1/15 fragment &0.351298 & 0.188586 & 0.135497 \\
 & 1/20 fragment &  0.310800 &0.163463 &0.162279 \\
  & whole genome &0.335605 &0.173103 &0.143191 \\
\hline
  & 1/4 fragment & 0.293139 &0.226877 &0.197907 \\
  & 1/8 fragment &0.275901 &0.220717 &0.206184 \\
C. trachomatis & 1/15 fragment & 0.299231 & 0.226269 & 0.194245 \\
  & 1/20 fragment & 0.293706 & 0.219299 &0.192447 \\
 & whole genome & 0.284452 & 0.223418 &0.201998 \\
\hline
  & 1/4 fragment & 0.253291 & 0.253147 &0.237551 \\
  & 1/8 fragment & 0.250753 &0.250494 &0.240300 \\
E. coli  & 1/15 fragment &0.256441 &0.248731 &0.232963 \\
  & 1/20 fragment &0.252115 &0.252027 &0.237276 \\
  & whole genome & 0.248986 & 0.255393 & 0.242893 \\
\hline
  & 1/4 fragment & 0.339263 & 0.165702 & 0.140649 \\
  & 1/8 fragment & 0.335415 & 0.187653 & 0.158851 \\
M. genitalium  & 1/15 fragment & 0.337408 & 0.173610 & 0.144801 \\
  & 1/20 fragment & 0.336145 & 0.182237 & 0.149540 \\
  & whole genome & 0.335212 & 0.175269 & 0.147534 \\
 \end{tabular}
 \end{center}
 \label{vectortab}     
\end{table}

\end{document}